\documentclass[12pt,preprint]{aastex}

\usepackage{natbib} 

\def\jpp{J. Plasma Phys.}
\def\phpl{Phys. Plasmas}
\def\ulysses{\emph{Ulysses}}
\def\k{{\bf k}}

\def\OmLuno{\Omega_{L1}}
\def\OmLdos{\Omega_{L2}}
\def\OmS{\Omega_{S}}
\def\kLuno{\k_{L1}}
\def\kLdos{\k_{L2}}
\def\modkLuno{k_{L1}}
\def\modkLdos{k_{L2}}
\def\kS{\k_{S}}
\def\OmL{\Omega_L}
\def\OmS{\Omega_S}
\def\kL{\k_L}
\def\wpe{\omega_{\rm pe}}
\def\VTe{V_{\rm Te}}
\def\lDe{\lambda_{\rm De}}
\def\cos#1{{\rm cos}\left(#1\right)}
\def\cosdos#1{{\rm cos}^2\left(#1\right)}
\def\NL{N^L}
\def\NS{N^S}
\def\VSW{{\bf V}_{\rm sw}}
\def\modVSW{V_{\rm sw}}
\def\tSr{ \theta_{\rm {\bf S} {\bf r} } }
\def\tLr{ \theta_{\rm {\bf L} {\bf r} } }

\def\tSL{\theta_{\rm \bf SL}}
\def\fpe{f_{\rm pe}}

\def\Vphi{V_{\phi}}
\def\deg{^{\circ}}
\def\Hz{{\rm Hz}}
\def\kHz{{\rm kHz}}
\def\cm{{\rm cm}}
\def\sec{{\rm seg}}
\def\K{{\rm K}}
\def\AU{{\rm AU}}
\def\mate#1{$#1$}

\def\corchetes#1{\left[#1\right]}
\def\parentesis#1{\left(#1\right)}

\def\eqn#1{equation (\ref{eq:#1})}

\def\Fig#1{Figure~\ref{fig:#1}}

\def\Dt{\Delta\theta}
\def\ez{{\bf e}_{\rm Z}}
\shorttitle{Electrostatic Decay of Plasma Turbulence}
\shortauthors{V\'asquez and G\'omez}

\begin{document}

\title{Electrostatic Decay of Beam-generated Plasma Turbulence}

\author{Alberto M. V\'asquez \altaffilmark{1} 
        and Daniel O. G\'omez\altaffilmark{1}}

\affil{Instituto de Astronom\'\i a y F\'\i sica del Espacio,\\
       CC 67 - Suc 28, (1428) Ciudad de Buenos Aires, Argentina.}       

\email{albert@iafe.uba.ar}

\altaffiltext{1}{
Also at the Department of Physics of the University of Buenos Aires, Argentina.}

\begin{abstract}

The study of the evolution of a suprathermal electron beam traveling through a 
background plasma is relevant for the physics of solar flares and their
associated type III solar radio bursts.
As they evolve guided by the coronal magnetic field-lines, these beams generate
Langmuir turbulence. The beam-generated turbulence is in turn 
responsible for the emission of radio photons at the second harmonic of the local 
plasma frequency, which are observed during type III solar radio bursts.
To generate the radio emission, the beam-aligned Langmuir waves must coalesce, and
therefore a process capable of re-directioning the turbulence in an effective fashion 
is required.
Different theoretical models identify the electrostatic (ES) decay process
\mate{L_1\rightarrow L_2+S}
(\mate{L}: Langmuir wave; \mate{S}: Ion-acoustic wave) 
as the re-directioning mechanism for the \mate{L} waves.
Two different regimes have been proposed to play a key role: 
the back-scattering and the diffusive (small angle) scattering.
This paper is a comparative analysis of the decay rate of the ES decay for
each regime, and of the different observable characteristics that are expected
for the resulting ion-acoustic waves.

\end{abstract}


\section{Introduction}

During solar flares, large amounts of energy are released and transformed in coronal 
heating and particle acceleration, where regions of magnetic reconnection are 
believed to be the acceleration
sites for suprathermal electron beams. Once accelerated, 
these beams travel through the coronal plasma guided by the coronal magnetic
field-lines, and generating a variety of observable emissions.
One long standing discrepancy in the modeling of this phenomena is that a 
beam with a given energy flux seems not to be able to simultaneously reproduce
the observed emissivities in HXR (due to non-thermal bremsstrahlung at the
chromosphere) and radio emission (due to beam-generated Langmuir turbulence in the
corona). Once the beam energy flux is set to reproduce the HXR levels, the 
derived radio emissivities tend to be much higher than observed levels 
\citep{ES84,HP87}.

In this context, we have developed a model for the evolution of electron beams
and the generation of Langmuir turbulence, and we have also computed 
the emission of radio waves due to the coalescence of the beam-excited Langmuir 
waves.
Our models describe the evolution of the beam and the produced Langmuir turbulence,
consistently considering the effect of both collisions and quasi-linear
wave-particle interaction. 
The level of turbulence derived from our models is up to two orders
of magnitude lower than previous attempts \citep{VG97}. 
The production of photons at the second harmonic
of the plasma frequency (radio waves), is the result of the coalescense
ot two Langmuir waves.
In our model we \emph{assume} that the beam-generated Langmuir waves become 
isotropic in an effective way, due to electrostatic decay
\mate{L_1+L_2\rightarrow T(2\wpe)}.
We found that an adequate treatment of the second harmonic photons 
generation (relaxing the head-on approximation) yields to further reductions
of the radio emission \citep{VGF02}.
Although these results help to reduce the gap between the predicted HXR and radio 
emissivities, we find that further reductions are required.

%MENCIONAR IMPORTANCIA DE MEDICIONES IN-SITU

%Nonetheless, an analysis
%of the characteristic time-scales involved in the process, indicates
%that the Langmuir wave re-directioning can occur fast enough to justify
%the isotropy assumption. In many cases though, the characteristic
%timescales of the isotropization and the beam-turbulence quasilinear interaction
%can be comparable, and a coupled treatment is required.
%A simultaneous consideration of this effect would favor a further reduction
%in the radio emission levels. This is due to the fact that its inclusion will imply
%a limitation of the effectiveness of the quasi-linear relaxation, yielding
%to lower Langmuir turbulence levels.

The observed radio emission requires the coalescense of the beam-generated Langmuir
waves. Therefore, a process capable of re-directioning the turbulence in an effective 
fashion is required. 
Different models in the literature resort to the electrostatic (ES) decay
\mate{L_1\rightarrow L_2+S} (\mate{L}: Langmuir wave; \mate{S}: Ion-acoustic wave) 
as the re-directioning mechanism for the \mate{L} waves.
Two different regimes have been proposed to play a key role.
One of them is the so called \emph{back-scattering} limit of the ES decay
\citep{ER01,C87}. In this
limit, the primary Langmuir wave decays into another one that propagates almost 
in the opposite direction.
The other asymptotic regime is the \emph{small-angle} limit
of the ES decay, which has also been pointed out by some authors 
as potentially relevant in this context \citep{M82,T70}. In this limit, 
the propagation directions of both Langmuir waves ($L_1$ and $L_2$) 
form a small angle, and the ES decay acts on the beam-generated Langmuir waves 
as a difussion mechanism through ${\bf k}$-space.
In our models described above,
the isotropization of Langmuir waves has been an assumption rather than a result
obtained from our beam-turbulence quasi-linear equations. 
If the timescale of the ES decay is short enough (as compared to the timescale
for the Langmuir waves to escape their generation region), it is reasonable to expect 
that the small-angle limit will render the beam-generated Langmuir turbulence isotropic.
%We have 
%favoured the small-angle (or diffusive) limit of the ES decay, and considered that, 
%as a result, the beam-generated Langmuir waves become isotropic \citep{VG03,VGF02}.
In this work we revise this approximation in detail. 
We compare the rate of ion-acoustic wave generation
in both limits, for a given set of beam-generated Langmuir spectra. We also analize
the resulting frequencies of the ion-acoustic
waves produced in each limit, and compare against reported in-situ observations.

In-situ observations of type III solar
radio bursts have shown clear evidence in support of the
ocurrence of the ES decay simultaneously with Langmuir waves.
For example, \citet{CR95} have analyzed ISEE 3 data that
shows the coexistence of high and low frequency electrostatic waves,
identified as Langmuir and ion-acoustic waves respectively,
during a type III radio event.
They analyze the observed low frequency waves, and find that their
frequencies are consistent with those predicted by assuming the 
existence of the ES decay 
acting in the back-scattering limit. Similar observational works
by \citet{T03} and \citet{TM98}, analyzing
URPE/\ulysses~ data, also support the occurence of the ES decay in 
association to impulsive Langmuir waves excited during type III radio events. 
In the present paper we make a similar analysis to the one by \citet{CR95}, and find out
that the URPE data analyzed by \citet{TM98} and \citet{T03} 
is consistent with the occurrence of the ES decay acting in the diffusive limit.

This paper is organized as follows. In section 2 we revisit the ES decay
and the general expression of its rate of occurrence. 
In section 3 we apply the general results to the specific case of
a beam-generated Langmuir spectrum. We make a quantitative comparison between 
the diffusive and back-scattering cases. In section 4 we compute the expected
frequencies for the generated ion-acustic waves for the particular type III solar
radio bursts observed and studied by \citet{TM98} and \citet{T03}. 
Finally, in section 5 we list our main conclusions.

\section{Electrostatic Decay Rate}

The electrostatic decay \mate{L_1\rightarrow L_2+S} (where
\mate{L_{1,2}}: Langmuir;  \mate{S}: Ion-acustic),
must verify the momentum and energy conservation, so that
their wave vectors and frequencies satisfy

\begin{equation}
{\kLuno = \kLdos + \kS} 
{\ \ \ ; \ \ \ }
{\OmLuno = \OmLdos + \OmS}
\end{equation}

\noindent 
where the dispersion relationships are,

\begin{equation}\label{eq:dispersion}
{\OmL(\kL)\approx  \wpe\parentesis{1+\frac{3}{2} (k_L \lDe)^2} }
{\ \ ; \ \ }
{\OmS(\kS)\approx  k_S V_S}
\end{equation}

\noindent
From these relations, we obtain the modulus and direction of the 
wave vectors of the resulting waves \mate{L_2} and \mate{S}
(see \Fig{sketch}),
with respect to the wave vector of the initial Langmuir wave
\mate{L_1}

\begin{eqnarray}
\cos{\alpha}&\equiv&\frac{\kLuno\cdot\kLdos}{k_{L1}k_{L2}}=\frac{\modkLuno-\mu \ k_S}{\modkLdos}\\ 
\mu\equiv\cos{\beta}&=&\frac{\kLuno\cdot\kS}{k_{L1}k_S} = \frac{k_S+2k_0}{2k_L} \ \ ; \ \
k_0\equiv\frac{1}{3} \frac{\omega_e}{V_{\rm Te}} \sqrt{\frac{m_e}{m_i}} \\
\modkLdos^2 &=& \modkLuno^2 - 2 k_0 k_S
\end{eqnarray}

The spectral density \mate{N^\sigma_\k} of the plasmons of type  
$\sigma$ (with $\sigma=L,~S$ in this case), is defined such that the volumetric density of
plasmons is \mate{n^\sigma = \int {d\k\over(2\pi)^3}\ N^\sigma_\k}.
In terms if these spectral densities,
the rate of production of ion-acoustic (S) waves is \citep{T70}

\begin{eqnarray}\label{eq:decayrate}
  \frac{d\NS_{\kS}}{dt} = \int {d\kLuno d\kLdos \over (2\pi)^6}
  w_L^{SL}  \times \nonumber \\
  \left[
  \NL_{\kLuno} \NS_{\kS}    +
  \NL_{\kLuno} \NL_{\kLdos} -
  \NL_{\kLdos} \NS_{\kS}
  \right]
\end{eqnarray}

\noindent where the probability of the decay is

\begin{eqnarray}
 w_L^{SL} = \frac{\hbar e^2\OmS^3 m_p (2\pi)^6}{8\pi m_e^3 V_{\rm Te}^4 k_S^2}
             \corchetes{\frac{\kLuno\cdot\kLdos}{\modkLuno\modkLdos}}^2 
              \delta\parentesis{\kLuno -\kLdos -\kS } 
              \times \nonumber \\
              \delta\parentesis{\OmLuno-\OmLdos-\OmS}
\end{eqnarray}

\noindent
where $m_e,~m_p$ are the electron and proton mass respectively,
$V_{\rm Te}$ is the electron thermal velocity,
and the delta functions express the momentum and energy conservation.

\section{Results for beam-generated Langmuir waves}

We use the expresions of the previous section to compute
the decay rate of a beam-generated Langmuir spectrum.
Hereafter we adopt a given shape for a beam generated
Langmuir spectrum. The dependence of the spectrum upon the
modulus of wavenumber, has been obtained by \citet{VG97}.
We also assume this spectrum to by axisymmetric and with
a gaussian angular spread about the direction of the beam,
and analyze the \emph{initial} ES decay rates.

The beam-generated Langmuir turbulence is initially aligned
with the beam propagation axis (the local magnetic field).
As it proceeds, we expect the ES decay to re-directionate the Langmuir 
waves. We thus divide our analysis in two cases.
In a first place we analyze the ES decay of a perfectly colimated
Lanmguir spectrum. Later on, we analize the same rates for
a non-colimated spectrum. In both stages of our analysis we
compute and compare the rates for both difussive and back-scattering cases.

\subsection{Colimated Langmuir spectra}

Let us evaluate the initial rate of production of S waves generated by 
a Langmuir spectrum, which is in turn produced by
a perfectly colimated beam (i.e. we take \mate{\NL_{\kLdos} = 0}),

\begin{eqnarray}\label{eq:rate}
  \frac{1}{\tau(\kS)}\equiv
  \frac{1}{\NS_{\kS}}\frac{d\NS_{\kS}}{dt} 
  \approx \int {d\kLuno d\kLdos \over (2\pi)^6}
  w_L^{SL} \NL_{\kLuno} = \nonumber \\ 
   \frac{K(T_e,n_e)}{(2\pi)^3} 
   \int {d\kLuno \over (2\pi)^3}\
  \frac{\corchetes{\kLuno\cdot(\kLuno-\kS)}^2}{\modkLuno^2(\kLuno-\kS)^2}\
  \NL_{\kLuno}\ \delta\parentesis{k_1 \mu - \frac{k_s+2k_0}{2} } 
\end{eqnarray}

\noindent
where the integral in \mate{\kLdos} has been performed using the momentum conservation,
and the function \mate{K} is given by

\begin{equation}\label{eq:K}
K(T_e,n_e)=\frac{\hbar e^2 \Omega_S^3 m_i (2\pi)^6 }{8\pi m_e^3 V_{\rm Te}^4k_S^2}\ 
           \frac{\wpe}{3 k_S V_{\rm Te}^2}
\end{equation}

\noindent
In a previous work \citep{VG97,VGF02} we developed a model to derive the
spectrum of beam-generated Langmuir waves.  According to this model,
the wavenumber of the excited Langmuir waves lays in the range

\begin{equation}\label{eq:range}
k_1 = [k_{\rm min},k_{\rm max}] \approx [6,12] \ k_0 \ \sqrt{T_6}
\end{equation}

\noindent
where \mate{T_6\equiv T_e [\K] /10^6}.
Approximating to a constant level for the perfectly colimated spectrum,
and calling \mate{\ez} to the direction of the beam,

\begin{equation}
 {\NL_{\kLuno} = (2\pi)^3 \frac{n^L}{\Delta k^L}
                           \delta(k_{1x}) \delta(k_{1y}) },
\end{equation}

\noindent
where \mate{\Delta k^L} is the spectral width, the result for the initial 
production rate of ion-acoustic waves is

\begin{eqnarray}\label{eq:tau0}
{\frac{1}{\tau(\kS)}=\frac{1}{\tau_0}\ P(k_S,\mu)} \\
{\frac{1}{\tau_0} \equiv \frac{\pi}{4} \frac{k_0}{\Delta k^L} 
                         \frac{\wpe W^L}{n_e m_e \VTe^2} } \nonumber
\end{eqnarray}

\noindent where the dimensionless function \mate{P(k_S,\mu)} is given by

\begin{eqnarray}\label{eq:proba}
 P(k_S,\mu)\equiv  \frac{1}{\mu}\ 
                   \frac{\corchetes{2+(k_S/k_0)(1-2\mu^2)}^2}
                   {4+(k_S/k_0)^2+4(k_S/k_0)(1-2\mu^2)}; ~~
{\rm if~~} 6\sqrt{T_6} < \frac{k+2k_0}{2\mu} < 12\sqrt{T_6}
\end{eqnarray}

\noindent
and is zero otherwise.
\Fig{P} shows the dimensionless decay rate 
$P(k_S,\mu)$ given by \eqn{proba}, 
as a function of the wave number $k_S$ and the 
angle \mate{\tSL} between \mate{\kS} and \mate{\kL}.
Here we have taken \mate{T_6=0.17} as a characteristic
number for a type III solar radio bursts analyzed in section 4. 
This graphic
confirms that the cases of maximum probability of occurrence are the
back-scattering (with \mate{\tSL\rightarrow 0,~ P\sim 1})
and the diffusive scattering (with \mate{\tSL\sim 75\deg,
~P\sim 4}). The factor of four in the diffusive decay rate
results from the modulation factor \mate{1/\mu} in \eqn{proba}.
%The right panel shows the probability as a function of the Cartesian components
%of the wave vector for the ion-acoustic wave: \mate{(k_{S(X)},k_{S(Y)})},
%where $X$ is the beam propagation direction (and the local magnetic field direction).
The propagation direction of the resulting ion-acoustic wave is
parallel to the beam propagation in the case of back-scattering, 
and almost perpendicular to the beam in the diffusive case (with \mate{k_S\ll k_L}).
We therefore find that the decay rate becomes maximum for two limiting cases:
a) small-angle (diffusive) scattering: \mate{\alpha\rightarrow 0}, and
b) back-scattering:     \mate{\alpha\rightarrow \pi}.

Under the diffusive scattering approximation, estimates for the characteristic
isotropization and energy transfer timescales, indicate
that the latter is much larger, implying that the isotropization 
occurs in a quasi-elastic fashion \citet{T70} \citep[see also][]{VGF02}. 
The quasi-conservation
of the total energy is readily seen from the following argument. The ES decay
conserves the total number of Langmuir plasmons. Now, given the weak dependence
of their energy on the wave-number (see \eqn{dispersion}), we can approximate 
the energy of each plasmon as \mate{\wpe}. 
Therefore, this decay can not produce a significant
change in the total Langmuir wave energy.

\subsection{Non-colimated Langmuir spectra}

In the previous section we found that the ES decay of the 
1D Langmuir spectrum generated by a perfectly colimated beam
is initially dominated by small-angle decay processes and, in second
place, by back-scattering.
We then expect that after a transient stage, the Langmuir spectrum is not
longer perfectly aligned with the beam direction, and that a growing
back-scattered spectrum also arises. Nonetheless, we expect that 
the forward spectrum will contain much more energy than the
backward spectrum, at least during the early stages of the evolution.
Under this assumption, we consider now an axisymmetric
spectrum around the beam direction, and model it through
functions \mate{N_{\bf k}=N(\theta,k)} that peak in
the forward direction and monotonically decrease as 
\mate{\theta\rightarrow\pi}.

Let us start by analyzing the small angle scattering limit, i.e. \mate{k_S \ll k_L}.
For the diffusive decay we have that the spectral width of the
ion-acoustic waves produced is much smaller than the spectral width of the primary
Langmuir waves. More specifically, we have

\begin{equation}
\frac{\Delta k_S}{\Delta k_L}\sim3\frac{\left<k_S\right>}{\left<k_L\right>}\ll 1
\end{equation}

\noindent
where \mate{\left<k_S\right>=k_{S{\rm max}}/2} and 
\mate{\left<k_L\right>=(k_{\rm min}+k_{\rm max})/2} indicate 
mean wavenumber values.
As the decay proceeds, each ES decay process will produce one ion-acoustic wave, hence

\begin{equation}
\NL \Delta k_L \sim \NS \Delta k_S  \rightarrow \frac{\NL}{\NS}\sim\frac{\Delta k_S}{\Delta k_L} \ll 1
\end{equation}

\noindent
i.e. the \emph{spectral} density of the (emitting) Langmuir waves becomes
much lower than that of the (emitted) ion-acoustic waves (even though the Langmuir energy 
density can be much larger than that of the ion-acoustic waves).  
Thus, we neglect the term \mate{\NL\NL} in the decay rate given by \eqn{decayrate}
and approximate

\begin{eqnarray}
  \frac{1}{\tau(\kS)}\equiv
  \frac{1}{\NS_{\kS}}\frac{d\NS_{\kS}}{dt} 
  \approx \int {d\kLuno d\kLdos \over (2\pi)^6}
  \ w_L^{SL} 
  \left( \NL_{\kLuno} - \NL_{\kLdos} \right) 
  = \\ \nonumber
  \frac{K}{(2\pi)^6} \int d\kLuno \frac{\cosdos{\alpha}}{\mu}
  \left( \NL_{\kLuno} - \NL_{\kLdos} \right) 
  \ \delta\left(k_{L1} -\frac{2k_0+k_S}{2\mu}\right) 
\end{eqnarray}

\noindent
where \mate{K} is given by \eqn{K} (see also \eqn{rate}).
Using the energy conservation, \eqn{rate1} reduces to the double
integral,

\begin{equation}\label{eq:rate1}
  \frac{1}{\tau(\kS)} \approx
  \frac{K}{(2\pi)^6} \int_0^{2\pi} d\phi_1 \int_0^\pi d\theta_1 
  \ \sin{\theta_1} \ 
  \left[
  \frac{k_1^2 \cosdos{\alpha}}{\mu} 
  \left( N(k_1,\theta_1) - N(k_2,\theta_2) \right)  
  \right]_{k_1=(k_S+2k_0)/(2\mu)}
\end{equation}

\noindent
where we have assumed the Langmuir spectrum \mate{N_{\bf k}} to be symmetric
about the beam pro\-pa\-ga\-tion axis \mate{\ez}
(i.e. independent of the angle \mate{\phi}),
and the values \mate{k_2,\theta_2} are related to \mate{k_1,\theta_1} through

\begin{eqnarray}
k_2 = \sqrt{k_1^2-2k_0 k_1}\\
\cos{\theta_2}=\frac{(\kLuno-\kS)\cdot\ez}{k_2}
\end{eqnarray}

\noindent
and \mate{\cos{\alpha}} is given by equation (3).
We note here that in the back-scattering limit, the term \mate{N_{L1} N_{L2}}
can not be neglected. Instead, for back-scattering, we expect that
\mate{N_{L2}\ll N_{L1}} (at early stages of the evolution),
and hence we can neglect the two last terms in \eqn{decayrate}. 
Therefore, the rate for back-scattering can be obtained from 
\eqn{rate1} by simply neglecting the second term \mate{N_{\theta_2 }},
and also taking the limits \mate{\cos{\alpha}\rightarrow -1}
and \mate{\mu\rightarrow +1}.

To compare these rates against the perfectly colimated case (\eqn{tau0}),
let us multiply and divide the RHS of \eqn{rate1}
by \mate{K_1 \equiv (2\pi)^3 n_L / \Delta k_L}, to obtain

\begin{equation}\label{eq:rate1b}
  \frac{1}{\tau(\kS)} =
  \frac{1}{\tau_0} P_1(k_S,\theta_S)
\end{equation}

\noindent 
where \mate{1/\tau_0} is the reference rate given by \eqn{tau0},
\mate{\theta_S} is the angle between the ion acoustic wave vector
and the beam propagation direction \mate{\ez},
and the dimensionless function \mate{P_1(k_S,\theta_S)} is given by

\begin{equation}\label{eq:P1}
  P_1(k_S,\theta_S) = \frac{\Delta k_L}{n_L} \frac{1}{(2\pi)^3} 
  \int_0^{2\pi} d\phi_1 \int_0^\pi d\theta_1 
  \ \sin{\theta} \ 
  \left[
  \frac{k_1^2 \cosdos{\alpha}}{\mu} 
  \left( N(k_1,\theta_1) - N(k_2,\theta_2) \right)  
  \right]_{k_1=(k_S+2k_0)/(2\mu)}
\end{equation}

We use this expression to evaluate the decay rate for a 
spectral model that is stronger in the forward direction (but not colimated). 
Here we neglect the dependence in wavenumber, and assume the spectral density
as constant within the range \mate{[k_{\rm min},k_{\rm max}]} 
given by \eqn{range}. More specifically we consider

\begin{equation}
N(k,\theta) = C \frac{1+e^{-\theta_c/\Dt}}{1+e^{(\theta-\theta_c)/\Dt}};
{\rm ~~if~~} k_{\rm min}<k<k_{\rm max}
\end{equation}

\noindent 
where the normalization constant \mate{C} is such that the number density of
Langmuir plasmons is \mate{n^L=\int\frac{{\bf k}}{(2\pi)^3} N_{\bf k}}.
As an example, \Fig{N} shows this function for 
\mate{\theta_c=25\deg,~\Delta\theta=2\deg}.
This angular distribution peaks at \mate{\theta=0} and monotonically decreases with 
increasing \mate{\theta}. The mean value is reached at \mate{\theta_c}.
Thus, for smaller values of this parameter the spectrum is more colimated.
The parameter \mate{\Dt} measures
the half-width over which the function varies from 75\% to 25\% of
its maximum value. For smaller values of this parameter, the spectrum
has a sharper angular edge.

\Fig{P1-05-25-02-diff} shows the resulting $P_1$ for two sharp-edged 
angular spectra with \mate{\Delta\theta=2\deg}. One spectrum is well concentrated
along the beam direction with \mate{\theta_c = 5\deg}, and the other
corresponds to a wider distribution with \mate{\theta_c = 25\deg}.
Our expression for $P_1$ (\eqn{P1}) is valid
in the diffusive regime \mate{k_S\ll k_L}, so we show results for
\mate{k_S = 0,.., \left<k_L\right>/10}.
Rates monotonically increase with \mate{k_S}, until they saturate at about
\mate{\left<k_L\right>/5}. For larger values of \mate{k_S} 
the rates decrease with increasing \mate{k_S}. 

Higher rates are obtained for a more colimated beam,
and the peak value is reached at a larger angle \mate{\theta_S}.
This behaviour can be understood in terms of our analytic decay analysis
for the perfectly colimated case. The peak is expected to occur 
as the result of the decay of Langmuir waves located in regions where the 
angular gradient of the spectrum is maximum, i.e. located around 
\mate{\theta_c}. This is because in this region the difference
\mate{N(k_1,\theta_1) - N(k_2,\theta_2)} in  \eqn{P1} reaches its largest value.
On the other hand, 
for the characteristic numbers of type III solar radio bursts 
being used in this analysis,
our results (see \Fig{P}) show that the ion-acoustic waves are 
preferentially directed toward angles around the value \mate{\theta_S \sim 75\deg}
(with respect to the direction of the decaying Langmuir wave).
We can therefore expect then that Langmuir waves with an angle \mate{\theta_c} respect
to the \mate{\ez} axis will preferentially
produce ion-acoustic waves with an angle \mate{\theta_S\sim 75\deg-\theta_c}.
This rough estimate gives peak angles of \mate{70\deg {\rm and}~50\deg} for
the two values of \mate{\theta_c} considered here, 
which is in agreement with the detailed numerical results shown in 
\Fig{P1-05-25-02-diff}. 

Also, in the perfectly colimated limit
(\mate{\theta_c\rightarrow 0, ~\Delta\theta \rightarrow 0}),
from comparison with \eqn{proba}, we expect that peak values of \mate{P_1}
will be close to the peak value \mate{P \sim 4} corresponding to the perfectly 
colimated case (see \Fig{P}). The numerical results of
\Fig{P1-05-25-02-diff} are consistent with this. As a consistency check,
\Fig{P1-0.5-0.5-diff} shows the same results for a highly colimated distribution with
\mate{\theta_c=0.5\deg,~ \Delta\theta=0.5\deg}, with peak rates
of \mate{P_1=3.5}, at \mate{\theta_S\sim 74\deg}.

For a fixed \mate{k_S=\left<k_L\right>/10},
\Fig{P1-05-25-02-10-diff} shows the effect of reducing the 
Langmuir wave distribution angular gradient. 
For the distribution with the edge centered on the
angle \mate{\theta_c=5\deg}, left panel shows the resulting rates for
\mate{\Delta\theta=2\deg,~10\deg}. The right panel shows the corresponding
results for the distribution with edge on the angle \mate{\theta_c=25\deg}.
The effect of a smoother distribution is to extend the possible angles
of the produced ion-acoustic waves over a much wider range.
Also, as the angular gradient of the Langmuir spectrum is reduced,
the difference \mate{N_1-N_2} in \eqn{P1} decreases,
and hence \mate{P_1(\theta_S)} yields lower values.

Let us also compare the different diffusive decay rates
against the corresponding back-scattering decay rates. 
The backscattering of a Langmuir plasmon of wavenumber \mate{k_L} produces
an ion-acoustic plasmon of wavenumber \mate{k_S = 2k_L-2k_0}, as
follows from taking the limit \mate{\mu\rightarrow+1} in equation (4).
\Fig{P1-05-25-02-back} shows the resulting back-scattering $P_1$ for the same two 
Langmuir spectra analyzed in \Fig{P1-05-25-02-diff}.
Thin curves correspond to \mate{k_S=2 k_{L,min}-2k_0}
and thick curves to \mate{k_S=2 k_{L,max}-2k_0}.
In this case, being the result of back-scattering, the angular distribution
of the decay rate closely follows that of the decaying Langmuir waves 
\mate{N(k_1,\theta_1)} in \eqn{P1}
(compare for example the right panel of \Fig{P1-05-25-02-back} with \Fig{N}).

A comparison between the back-scattering results of
\Fig{P1-05-25-02-back} against the diffusive decay results \Fig{P1-05-25-02-diff} 
shows that the diffusive rates are always larger, by factors of order 2 to 5.
\Fig{P1-0.5-0.5-back} shows the same results for a highly colimated distribution with
\mate{\theta_c=0.5\deg,~ \Delta\theta=0.5\deg}. The resulting rates are 
of order \mate{P_1\sim 1.5}, or about a factor of 2 lower
than the corresponding diffusive decay values of \Fig{P1-0.5-0.5-diff}.
These results are consistent with the perfectly colimated beam results of \Fig{P}.
Therefore we find that, in all cases,
diffusive decay rates are systematically larger
(though comparable) than the corresponding back-scattering rates. 

\section{Frequencies of the ion-acoustic Waves}

Several space-based missions, located at distances of order
$1~\AU$ from the Sun, are designed to perform in-situ measurements
of the extended solar wind. These missions are able then to perform in-situ
measurements of type III solar radio bursts, that sometimes extend far away 
from the Sun and reach the Earth. One of these instruments is 
the \emph{Unified Radio and Plasma Wave Experiment}
(URAP) aboard the \ulysses~ mission. For the purpose of
interpreting type III burst data collected by this experiment, let us suppose
that an ion-acoustic wave travels through the region of the experiment.
If \mate{\VSW} is the local solar wind velocity, the Doppler-shifted frequency 
of the ion-acoustic wave, as measured by the experiment, is given by

\begin{eqnarray}\label{eq:fs}
f_S &=& \frac{1}{2\pi}
            \parentesis{ k_S V_S + \kS \cdot \VSW } = \nonumber \\
            &=& \frac{k_S}{2\pi} \parentesis{V_S+\modVSW \ \cos{\tSr} } = \nonumber \\
            &=& 2 \frac{\fpe}{\Vphi}
            \parentesis{\cos{\tSL}-\frac{V_S \Vphi}{3\VTe^2}}
            \parentesis{V_S+\modVSW \ \cos{\tSr}}
\end{eqnarray}

\noindent
where $k_S$ has been eliminated from equation (4),
\mate{\tSL} and \mate{\tSr} are the angles between the propagation direction 
of the ion-acoustic wave and those of the primary (beam aligned)
Langmuir waves, and the (radial) solar wind, respectively. 
This formula is analogous to the one used by \citet{CR95}
to analyze a type III solar radio burst. They assume the ES decay products
only in the back-scattering limit, and hence they consider the particular case 
\mate{\cos{\tSL} = +1}. Also in that limit, the produced ion-acoustic
waves propagate aligned with the beam direction, so that \mate{\tSr=\tLr}, i.e.
the angle between the primary Langmuir waves and the
solar wind direction.
On the other hand, in the diffusive limit, 
\mate{\tSL>0} so that the angle \mate{\tSr} takes the range of values 
\mate{|\tSL-\tLr|<\tSr <\tSL+\tLr}.
% In these expresions
%\mate{\tLr} is the angle between the local directions of the solar 
%wind and the primary Langmuir wave propagation propagation direction.

To quantify the predicted frequencies for the ion-acoustic waves in each of these
two limiting cases, we refer to a specific observational case,
analyzed by \citet{TM98}. They show the data of a type III solar radio event
recorded by URAP/\ulysses~ on March 14 1995.
Their detailed analysis shows radio emissions at both the plasma and second harmonic
frequencies. At the same time, impulsive
electrostatic fluctuations were recorded at the local plasma frequency, which turns out
to be \mate{\fpe \sim 23~\kHz}, identified as Langmuir wave bursts.
The WFA (Wave Form Analyzer) instrument detected electric fluctuations,
highly correlated in time with the observed Langmuir impulsive peaks,
in the range $0\rightarrow 450\ \Hz$.
The instrument also detected magnetic field fluctuations in the range
$0\rightarrow 50\ \Hz$. The combination of these facts indicates
that the electric field fluctuations in the range $50\rightarrow 450 \ \Hz$ 
are of a pure electrostatic nature, such as ion-acoustic waves.
If this is the case, its strong temporal correlation with the Langmuir bursts 
supports the idea that the ion-acoustic 
waves may be produced by the ES decay of the L-waves \citep[see also][]{CR95}.
On the other hand, below $50\ \Hz$ fluctuations are most likely
of electromagnetic nature, such as whistlers.
Summarizing, the analysis by \citet{TM98} strongly suggests the presence of
ion-acoustic waves in the range $50\rightarrow 450 \ \Hz$, for this particular
event, and that these waves are produced as the result of the 
ES decay of the beam-excited Langmuir waves.

For the event under consideration, the numerical values of the
relevant parameters are \citep{TM98}:
     \mate{V_{\rm S}  =5.1\times 10^6\cm/\sec}, 
     \mate{\VTe =1.6\times 10^8\cm/\sec} (corresponding to $T_e=1.7\times10^5\K$),
     \mate{V_{\rm sw} =3.4\times 10^7\cm/\sec},
     \mate{\Vphi \sim V_{\rm beam}\ \pm \ 25\%},
     with a mean beam velocity estimated to be \mate{V_{\rm beam} = 3.5\times 10^9\cm/\sec},  
     \mate{\fpe  =2.3\times 10^4\Hz},
     \mate{\tLr\sim 45\deg}.

Using these numerical values we compute now, in the two limit 
cases under analysis, the predicted ion-acoustic wave frequencies.
In all cases, we assume that the primary (beam-excited) Langmuir waves form a mean 
angle $\tLr=45\deg$ with respect to the solar wind direction,
and that the primary Langmuir waves have phase
speeds in the range $V_\phi = V_{\rm beam} \pm 25\%$.

\Fig{P} shows that, in the diffusive limit, decays proceed
most likely for angles $\tSL$ in the range $60\deg \rightarrow 75\deg$ (in this range the
probability is about twice the back-scattering probability).
Using this range of angles in \eqn{fs},
we find that the diffusive-scattering assumption yields ion-acoustic waves
in the range $50\ \Hz \rightarrow 215 \Hz$.
On the other hand, if back-scattering is assumed, we have $\tSL=0\deg$, and
\eqn{fs} yields ion-acoustic frequencies in the range $215\ \Hz \rightarrow 415\ \Hz$.
The difference between the predictions in both limits 
is due to the term \mate{\parentesis{\cos{\tSL}-\frac{V_S \Vphi}{3\VTe^2}}}, 
that is maximized in the back-scattering case as \mate{\cos{\tSL}\sim +1}.

If the the primary Langmuir waves form a variable angle
$0\deg < \tLr < 90\deg$ with respect to the solar wind direction,
the frequency ranges predicted in both limits overlap each other,
within the observed range. In this case,
the back-scattering limit yields predicted ion-acoustic frequencies over the
whole observed range \citep[as already pointed out by][]{TM98}.
We thus find that the predictions, for both asymptotic cases are
consistent with the observations. Taking mean values of
the different parameters, we find out that
for the back-scattering assumption the predicted frequencies
are consistent with the higher frequency portion of the observations
\citep[see also][]{CR95}.
On the other hand, the diffusive-scattering limit yields predicted 
frequencies that are consistent with the lower frequency portion
of the observed electrostatic fluctuations.

Further empirical support arises from a recent observational work by
\citet{T03}, where they analyze URAP data for another type III
burst, in a similar fashion to the case already described. This case
corresponds to in-situ observations at much larger heliocentric distances,
specifically at \mate{5.2~\AU}, where the local ambient plasma parameters are 
very different ($\fpe \sim 2~ \kHz,~T_e\sim8.8\times10^4\K$) to those of the
previous case. The beam velocity (and hence the primary Langmuir waves
phase velocity) is estimated to be
\mate{\Vphi \sim 1.5\times 10^9\cm/\sec \ \pm \ 30\%}.
In this other case, the WFA-URAP instrument also detected electric field
fluctuations correlated in time with Langmuir bursts.
The electric waves that can be
safely assumed to be of electrostatic nature (i.e. with no simultaneous
detection of magnetic field fluctuations above the background level)
are in the frequency range $20\rightarrow 200\ \Hz$.

We repeated our probability calculations for this case,
and find out that, in the diffusive limit, decays proceed
most likely for angles $\tSL$ in the range $55\deg \rightarrow 73\deg$ 
(in this range the
probability is about twice the back-scattering probability).
Using this range of angles,
in the diffusive limit, the predicted ion-acoustic frequencies result
in the range $0\Hz \ \rightarrow 60\ \Hz$. On the other hand, under the
back-scattering assumption the predicted frequencies are in the range
$55\ \Hz \rightarrow 105 \ \Hz$.
%We thus find a similar situation as in the previous event:
%the back-scattering assumption is able to predict the higher frequency
%portion of the observations, while the diffusive-scattering limit
%is consistent with the lower frequency portion.

\section{Conclusions}

From the theoretical point of view, an analysis of the ES decay rate
indicates that the process is dominated by two limiting cases: diffusive
(small angle) scattering and back-scattering. We find that the decay
rates are comparable for both cases, being the diffusive 
decay rates systematically larger than those of the back-scattering limit.
Given the results of our quantitative
comparative analysis, we believe that both limiting cases are
acting at the same time. Furthermore, we speculate the net effect 
of diffusive decay and back-scattering acting simultaneously is
then that of a diffusive isotropization of the beam-generated Langmuir spectrum.
The back-scattering generates a backwards directed spectrum 
(respect to the beam propagation direction), but then this
secondary spectrum can also diffuse by small-angle decay.
Thus, if the ES decay is assumed to be present in solar flare and type
III solar radio burst scenarios, we believe that its capability to isotropize
Langmuir waves can not be neglected.

From an observational point of view, we refer to analyses of a type III
bursts by \citet{TM98} and \citet{T03}.
They registered low frequency electrostatic fluctuations
which are strongly correlated in time with impulsive 
Langmuir wave bursts. We find that the observed frequencies of
these fluctuations are consistent with the predicted
frequencies of ion-acoustic
waves produced by the ES decay of the primary (beam-excited) Langmuir waves.
Under the back-scattering assumption the predicted frequencies
are consistent with the higher frequency portion of the observations.
On the other hand, the diffusive-scattering limit yields predicted 
frequencies that are consistent with the lower frequency portion
of the observed electrostatic fluctuations.
The relative burst intensities at the two frequency ranges 
could then serve as a proxy for the relative effectiveness of the occurrence 
of the ES decay in both limits, that
we anticipate to be comparable from our theoretical analysis.

From this analysis, we speculate that both the back-scattering and diffusive limits
of the ES decay may be relevant in type III bursts (and presumably also in solar 
flare radio events). 
Also, the beam-generated Langmuir turbulence may become isotropic as the
ES decay proceeds mainly in the two limit cases of diffusive scattering and 
back-scattering. We postpone for a future work 
the investigation of the coupling between the ES decay and the quasi-linear
beam-turbulence interaction. We believe that the simultaneous
consideration of both effects can bring HXR and radio emission predictions into
a better agreement with observations.
This is due to the fact that its inclusion will imply
a limitation of the effectiveness of the quasi-linear relaxation, yielding
to lower Langmuir turbulence levels, and hence lower radio emissivity.

\acknowledgments 

We thank the anonymous referee for useful suggestions that helped to clarify
our main conclusions.
This work was funded by the \emph{Agencia Nacional de Promoci\'on de Ciencia y 
Tecnolog\'\i a} (ANPCyT, Argentina) through grant 03-09483.
We also thank \emph{Fundaci\'on Antorchas}
(Argentina) for partial support through grant 14056-20.

\clearpage

%
% Figure captions:
%

\figcaption[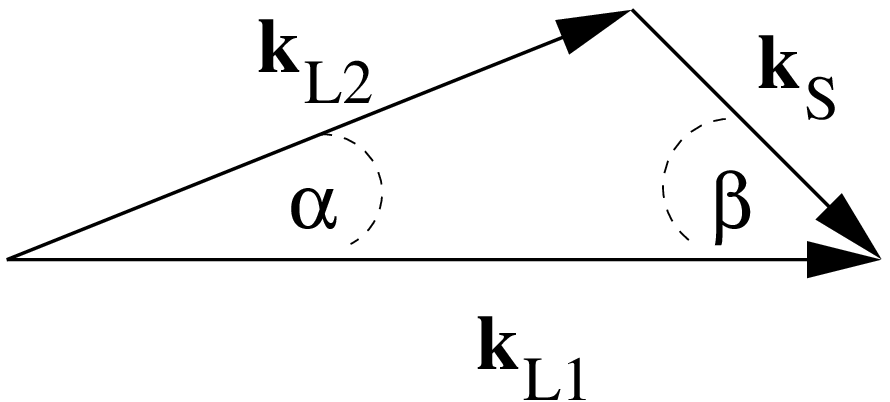]
{Momentum conservation in the electrostatic decay,
${\bf k}_{L1} = {\bf k}_{L2} + {\bf k}_S $. Angles $\alpha$
and $\beta$ are given by equations (4) and (5).
\label{fig:sketch}}

\figcaption[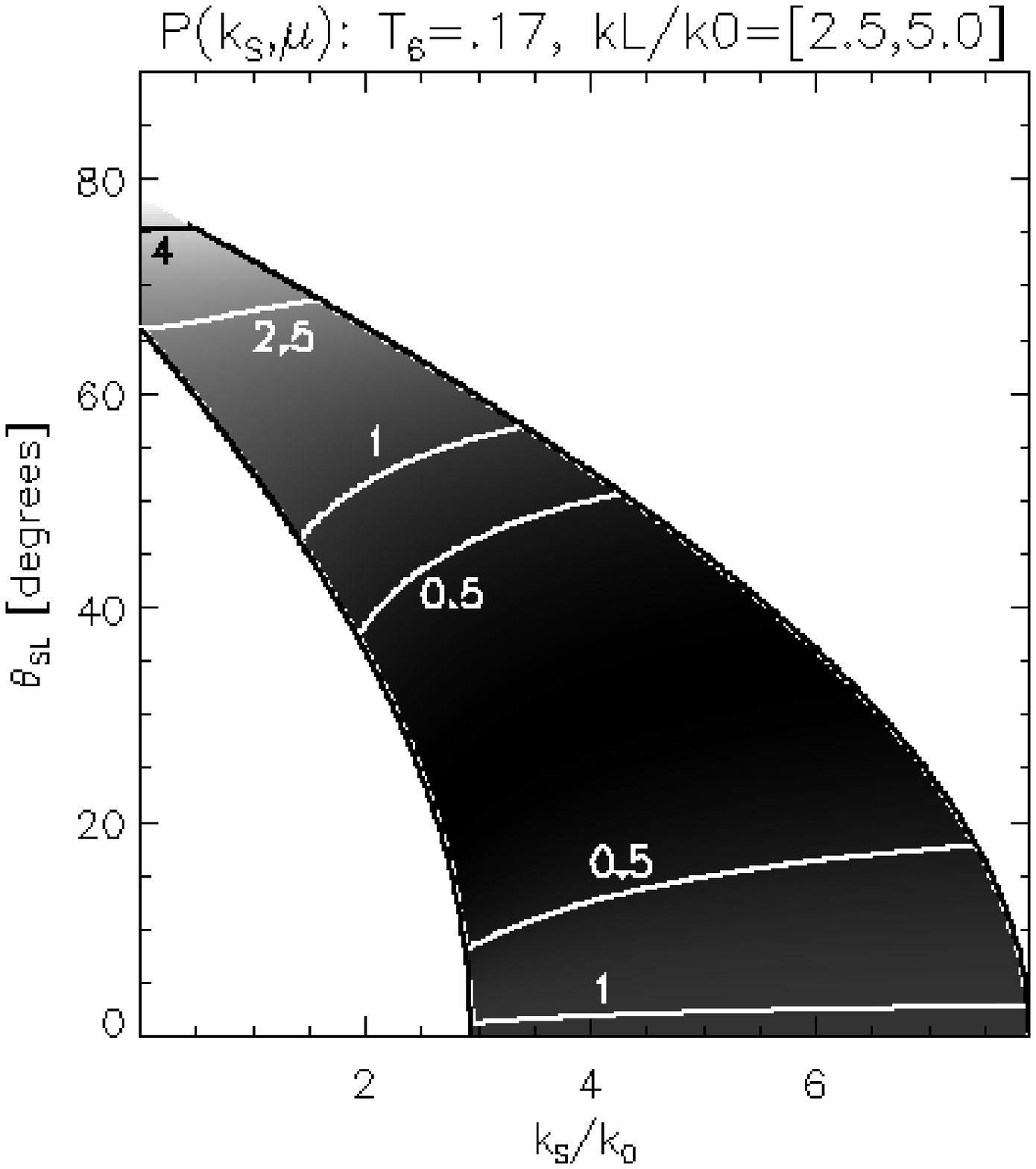]
{Dimensionless decay rate \mate{P(\kS)}, 
as a function of \mate{(k_S,\tSL)}.
\label{fig:P}}

\figcaption[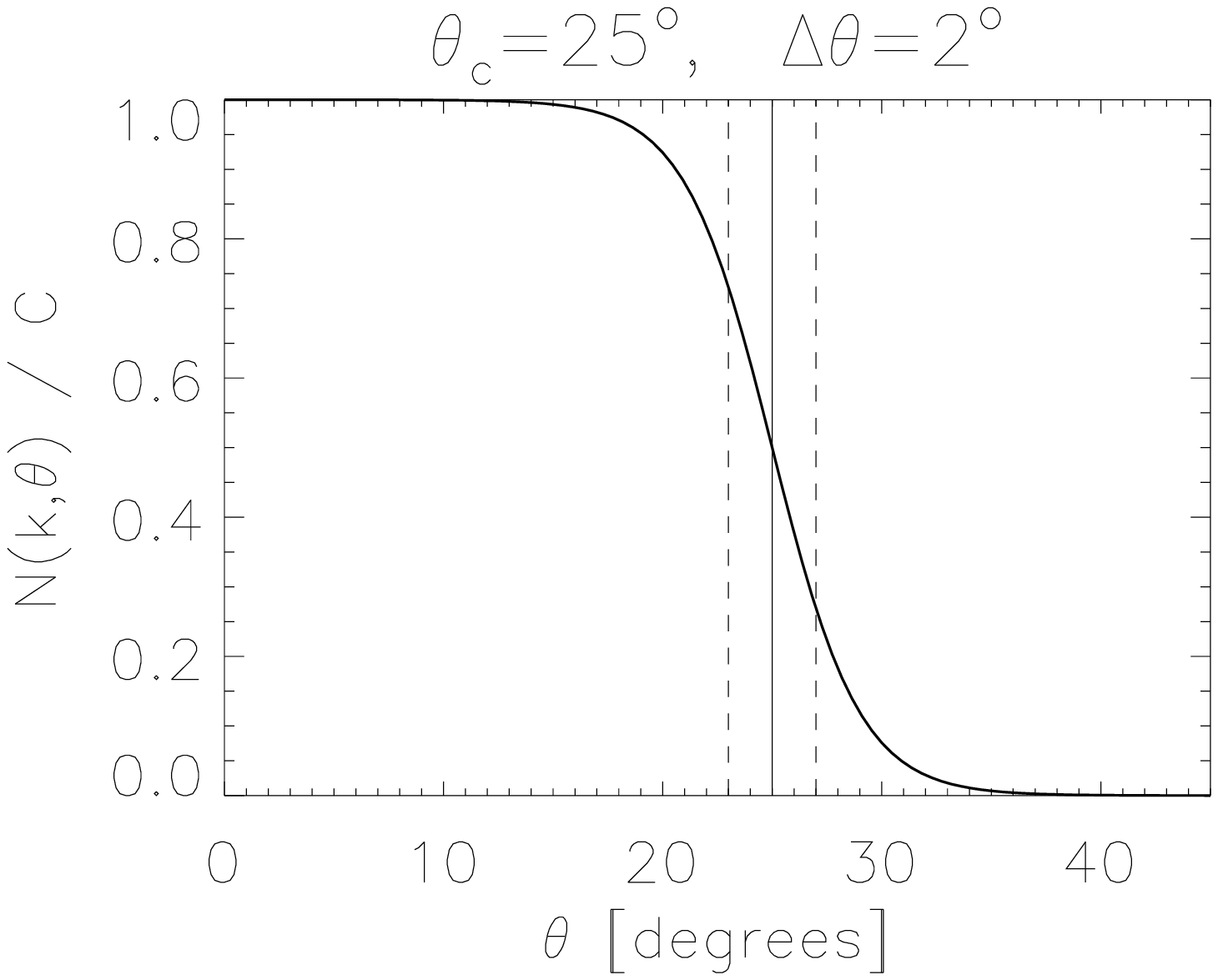]
{Model for the angular distribution of the Langmuir spectrum,
with \mate{\theta_c = 25 \deg} and \mate{\Delta\theta=2\deg}.
The vertical lines indicate the angles \mate{\theta_c} (full line),
and \mate{\theta_c\pm\Delta\theta} (dashed lines).
\label{fig:N}}

\figcaption[f4.eps]
{Diffusive decay: dimensionless factor \mate{P_1(k_s,\theta_S)} for 
\mate{\theta_c = 5\deg,~ 25 \deg}, and \mate{\Delta\theta=2\deg}.
Profiles are ploted as a function of \mate{\theta_S},
with different curves corresponding to \mate{k_S=0,..,\left<k_L\right>/10}
(curves with larger $P_1$ values correspond to larger \mate{k_S)}.
\label{fig:P1-05-25-02-diff}}

\figcaption[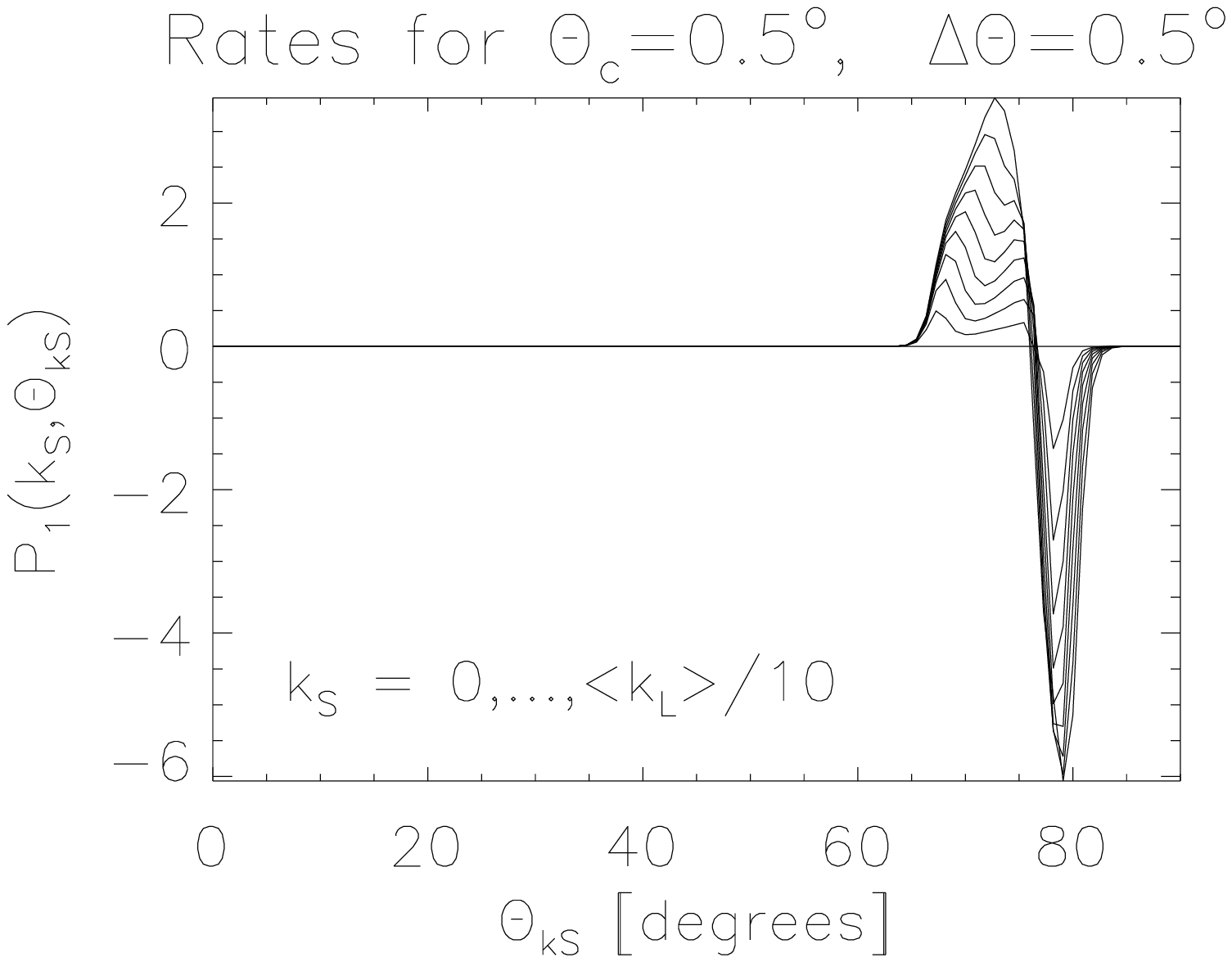]
{Diffusive decay: dimensionless factor \mate{P_1(k_s,\theta_S)} for 
\mate{\theta_c = 0.5\deg} and \mate{\Delta\theta=0.5\deg}.
Profiles are ploted as a function of \mate{\theta_S},
with different curves corresponding to \mate{k_S=0,..,\left<k_L\right>/10}
(curves with larger $P_1$ values correspond to larger \mate{k_S)}.
\label{fig:P1-0.5-0.5-diff}}

\figcaption[f6.eps]
{Diffusive decay: for a fixed \mate{k_S=\left<k_L\right>/10},
dimensionless factor \mate{P_1(k_s,\theta_S)} for 
\mate{\theta_c = 5\deg,~25 \deg}, and \mate{\Delta\theta=2\deg,~ 10\deg}.
Profiles are ploted as a function of \mate{\theta_S},
with thin curve corresponding to \mate{\Delta\theta=2\deg}
and thick curve to \mate{\Delta\theta=10\deg}.
\label{fig:P1-05-25-02-10-diff}}

\figcaption[f7.eps]
{Back-scattering: dimensionless factor \mate{P_1(k_s,\theta_S)} for 
\mate{\theta_c = 5\deg,~ 25 \deg}, and \mate{\Delta\theta=2\deg}.
Profiles are ploted as a function of \mate{\theta_S},
with thin curves corresponding to \mate{k_S=2 k_{L,min}-2k_0}
and thick curves to \mate{k_S=2 k_{L,max}-2k_0}.
\label{fig:P1-05-25-02-back}}

\figcaption[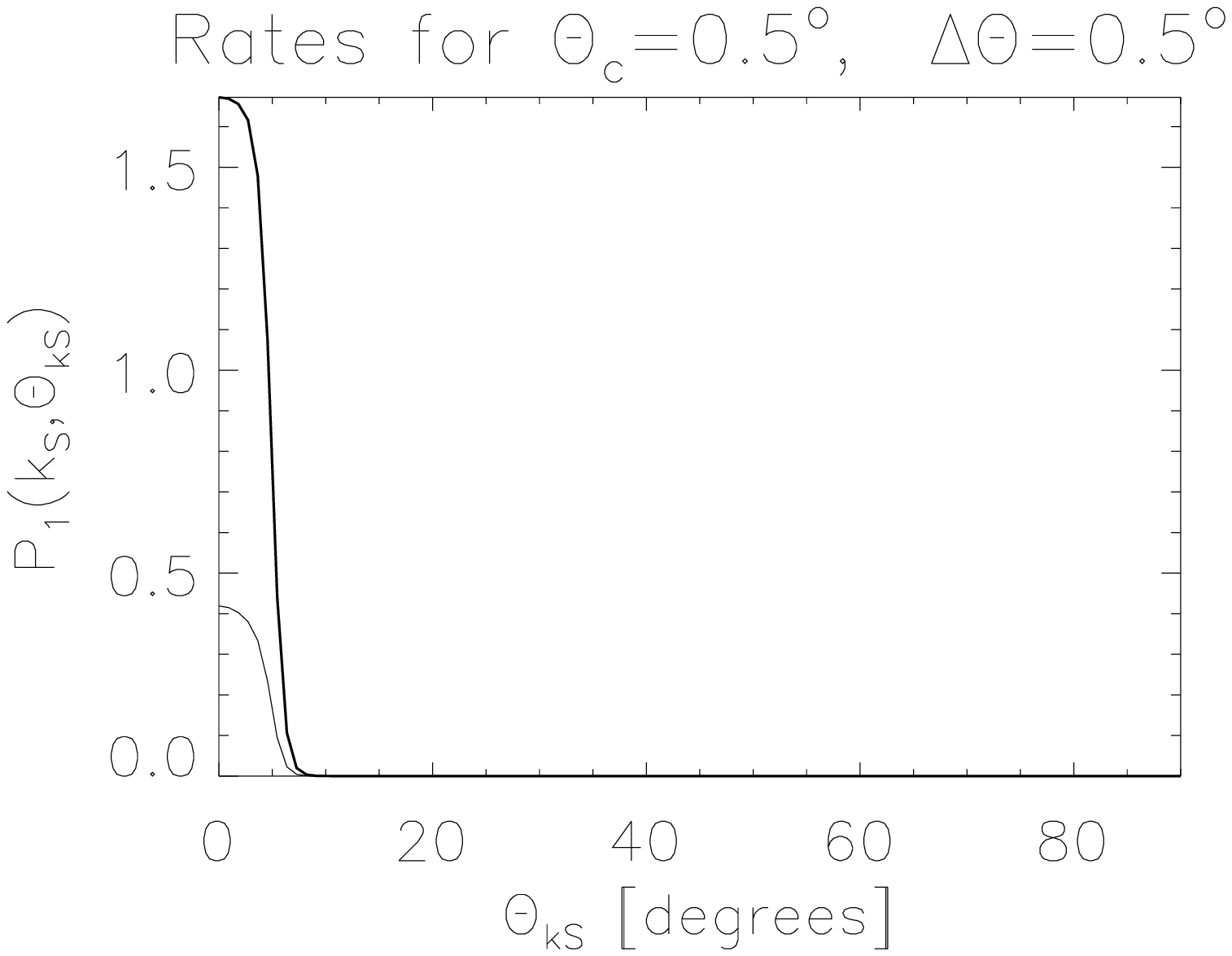]
{Back-scattering: dimensionless factor \mate{P_1(k_s,\theta_S)} for 
\mate{\theta_c = 0.5\deg} and \mate{\Delta\theta=0.5\deg}.
Profiles are ploted as a function of \mate{\theta_S},
with thin curves corresponding to \mate{k_S=2 k_{L,min}-2k_0}
and thick curves to \mate{k_S=2 k_{L,max}-2k_0}.
\label{fig:P1-0.5-0.5-back}}

\clearpage

%
% Figures:
%

\begin{figure}
\figurenum{1}
\epsscale{.5}
\plotone{f1.eps}
\caption{}
\end{figure}

\begin{figure}
\figurenum{2}
\epsscale{.75}
\plotone{f2.eps}
\caption{}
\end{figure}

\begin{figure}
\figurenum{3}
\epsscale{0.5}
\plotone{f3.eps}
\caption{}
\end{figure}

\begin{figure}
\figurenum{4}
\epsscale{1}
\plottwo{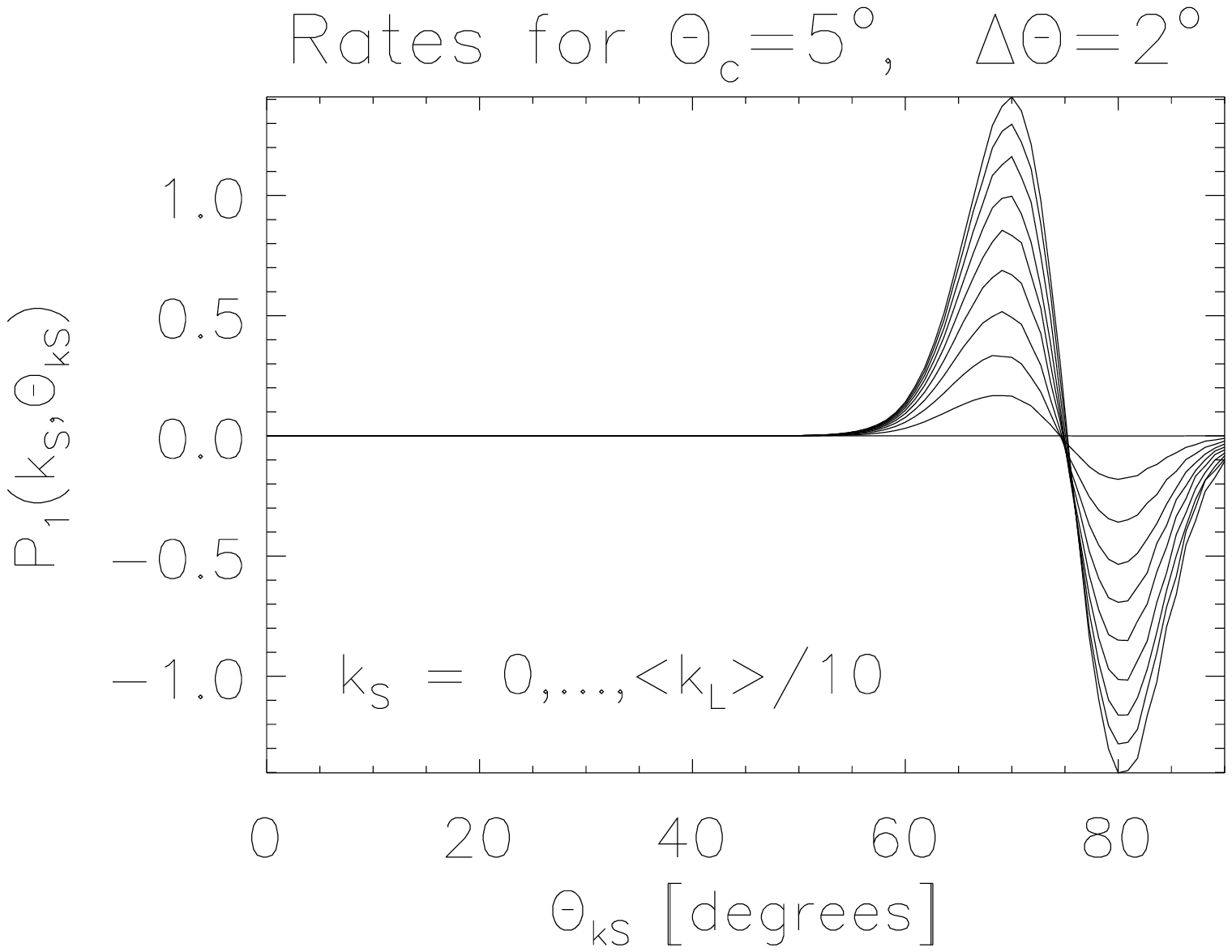}{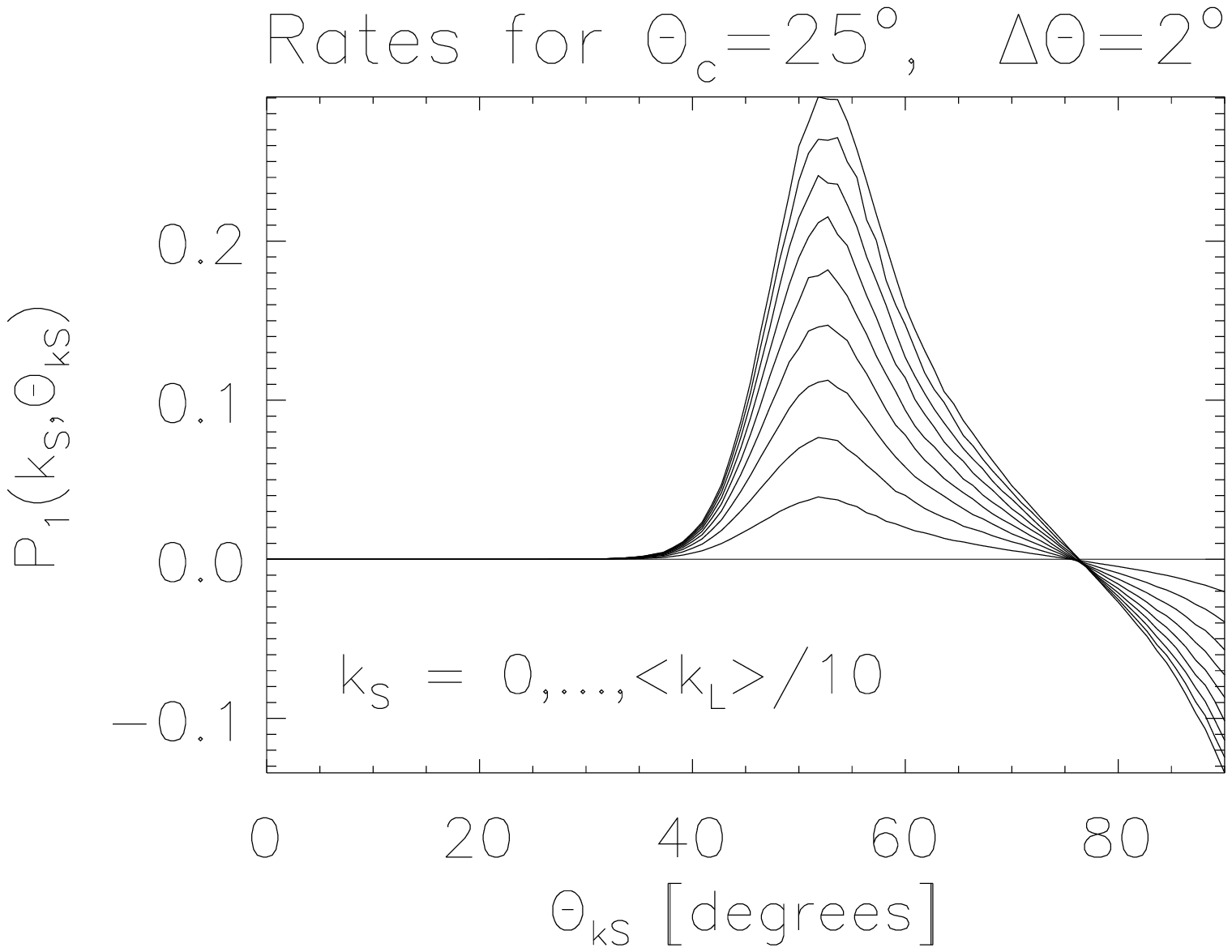}
\caption{}
\end{figure}

\begin{figure}
\figurenum{5}
\epsscale{0.5}
\plotone{f5.eps}
\caption{}
\end{figure}

\begin{figure}
\figurenum{6}
\epsscale{1}
\plottwo{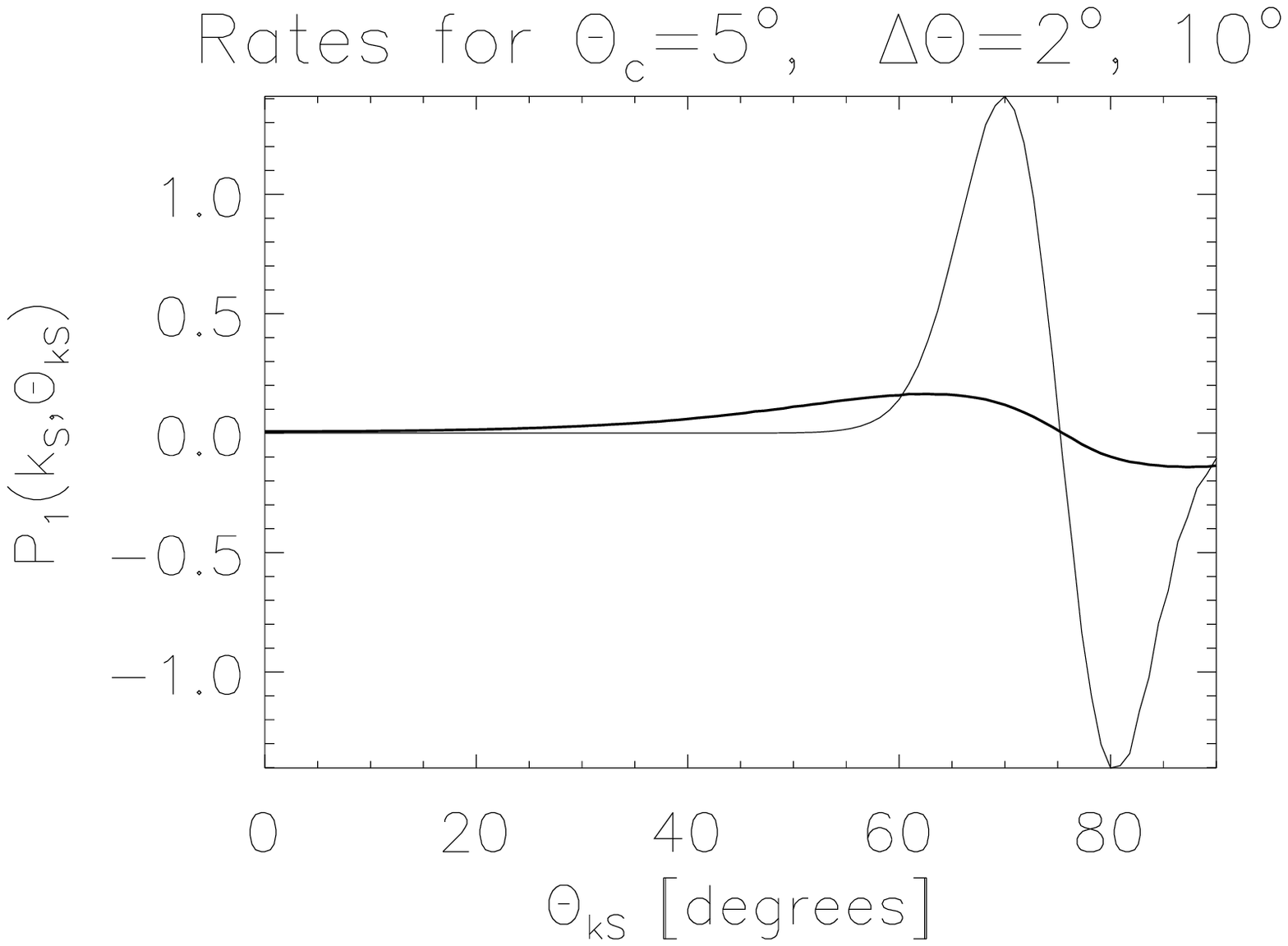}{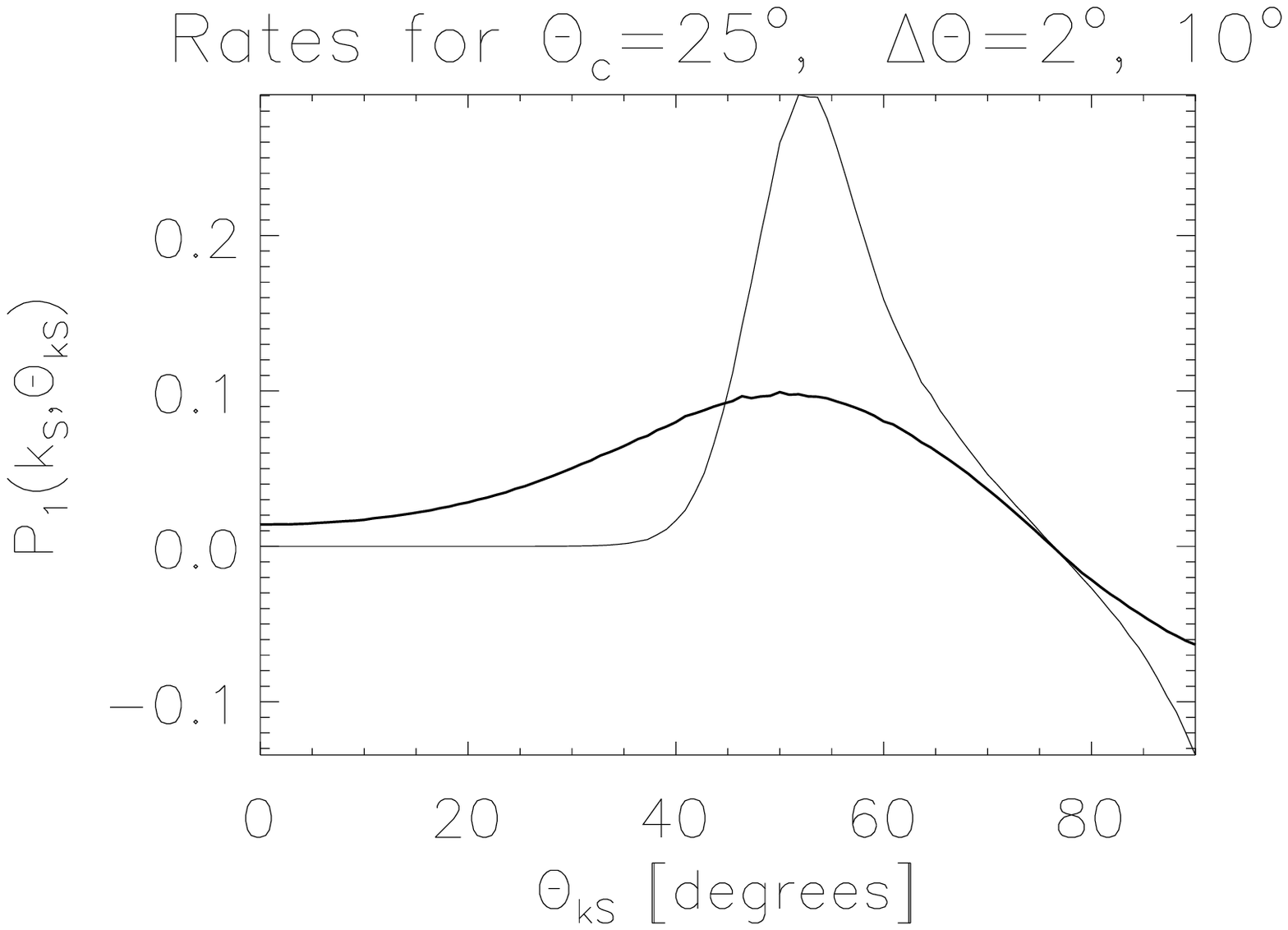}
\caption{}
\end{figure}

\begin{figure}
\figurenum{7}
\epsscale{1}
\plottwo{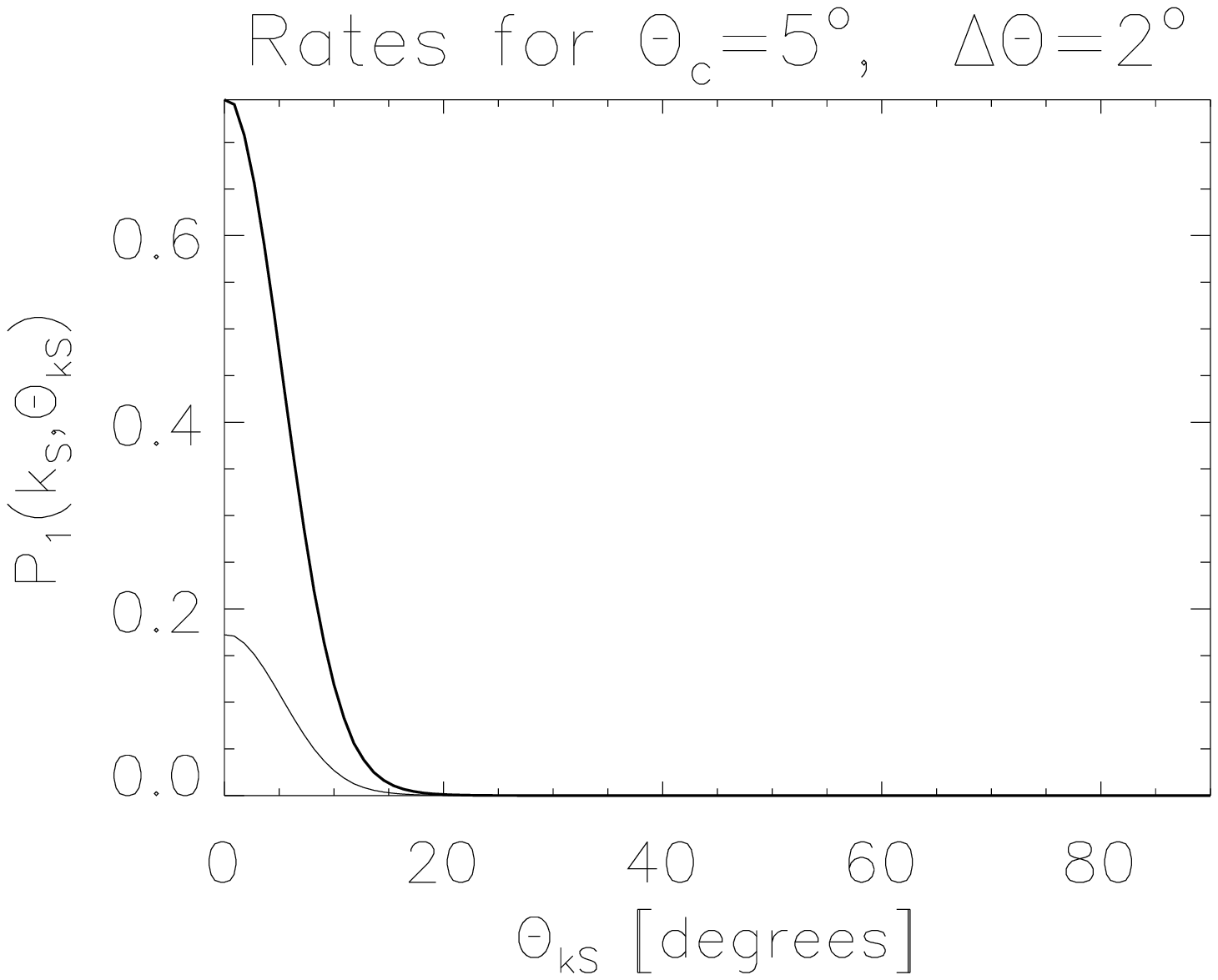}{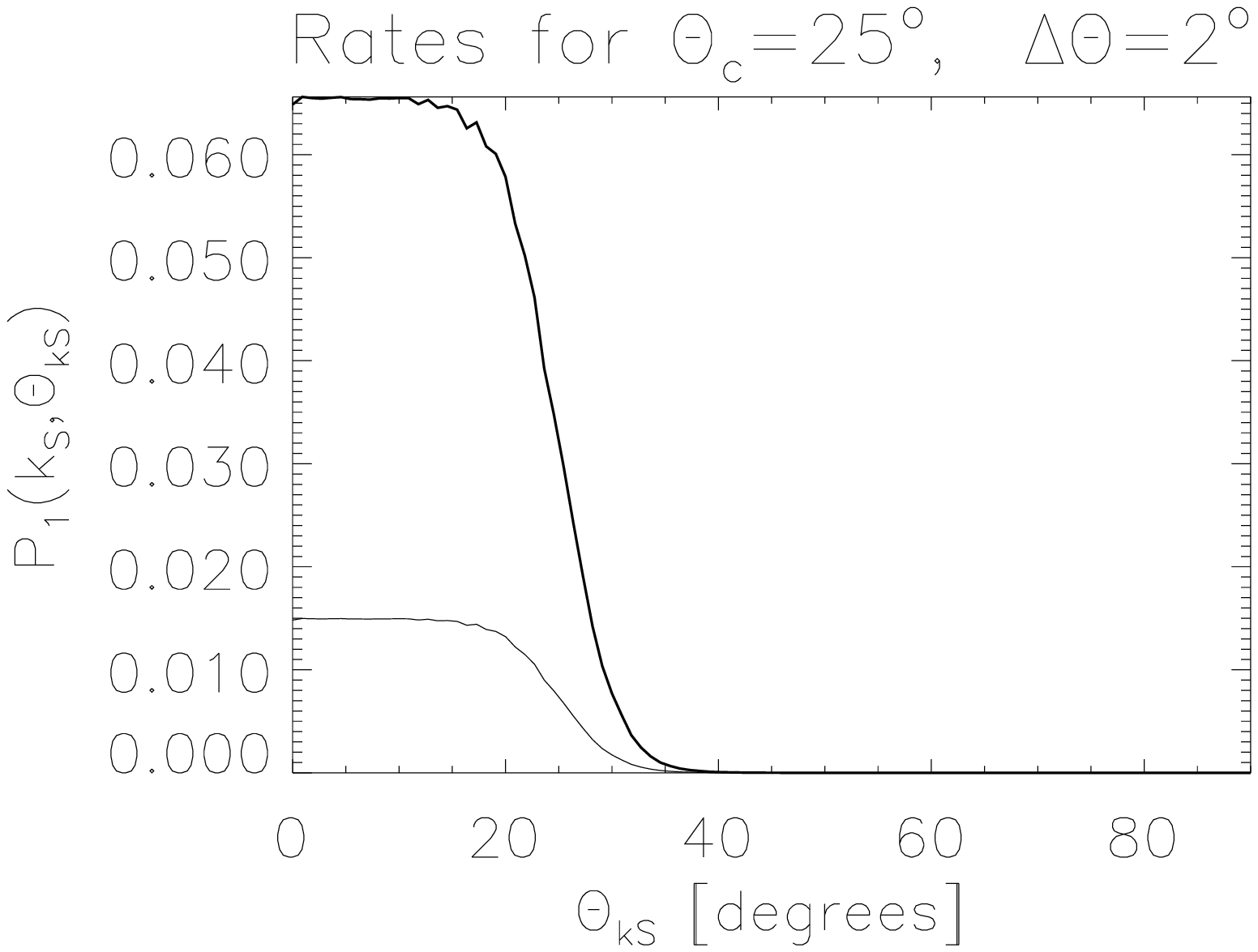}
\caption{}
\end{figure}

\begin{figure}
\figurenum{8}
\epsscale{0.5}
\plotone{f8.eps}
\caption{}
\end{figure}

\end{document}